\documentclass[aip,jcp,amsmath,amssymb,reprint,superscriptaddress]{revtex4-1}
\usepackage{graphicx}%
\usepackage{dcolumn}%
\usepackage{bm}%
\usepackage{epstopdf}
\usepackage[hyperindex,breaklinks]{hyperref}
\usepackage[caption=false]{subfig}

\newcommand{\tlc}{\tau_{LC}}
\newcommand{\tvh}{\tau_{VH}}
\newcommand{\zp}{\text{Zr}_{80}\text{Pt}_{20}}

\begin{document}
\title{Experimental determination of the temperature-dependent Van Hove function in a $\zp$ liquid}
\date{\today}
\author{R. Ashcraft}
\affiliation{Department of Physics, Washington University, St. Louis, Missouri, 63130, USA}
\author{Z. Wang}
\affiliation{Department of Materials Science and Engineering, Department of Physics and Astronomy, University of Tennessee, Knoxville, Tennessee 37996, USA}
\author{D. L. Abernathy}
\affiliation{Neutron Scattering Division, Oak Ridge National Laboratory, Oak Ridge, Tennessee 37831, USA}
\author{D. G. Quirinale}
\affiliation{Neutron Technologies Division, Oak Ridge National Laboratory, Oak Ridge, Tennessee 37831, USA}
\affiliation{Department of Physics and Astronomy, Iowa State University, Ames, Iowa 50011, USA}
\author{T. Egami}
\affiliation{Department of Materials Science and Engineering, Department of Physics and Astronomy, University of Tennessee, Knoxville, Tennessee 37996, USA}
\author{K. F. Kelton}
\email[Author to whom correspondence should be addressed: ]{kfk@wustl.edu}
\affiliation{Department of Physics, Washington University, St. Louis, Missouri, 63130, USA}
\affiliation{Institute of Materials Science and Engineering, Washington University, St. Louis, Missouri, 63130, USA}
\begin{abstract}
Even though the viscosity is one of the most fundamental properties of liquids, the connection with the atomic structure of the liquid has proven elusive.  By combining inelastic neutron scattering with the electrostatic levitation technique the time-dependent pair-distribution function (\emph{i.e.} the Van Hove function) has been determined for liquid $\zp$.  We show that the decay-time of the first peak of the Van Hove function is directly related to the Maxwell relaxation time of the liquid, which is proportional to the shear viscosity.  This result demonstrates that the local dynamics for increasing or decreasing the coordination number of local clusters by one determines the viscosity at high temperature, supporting earlier predictions from molecular dynamics simulations.  
\end{abstract}
\maketitle

\section{Introduction}\label{intro}
	The viscosity, $\eta$, of liquids shows common behavior among various disparate groups of liquids~\cite{Angell1995a}.  At high temperatures it has an Arrhenius temperature dependence with a constant activation energy.  But below a certain temperature, the viscosity crossover temperature, $T_A$, it becomes super-Arrhenius.  Kivelson~\cite{Kivelson1996} first showed that the viscosity of various liquids can be scaled onto one curve as a function of $T/T_A$ ($T_A$ was designated as $T^*$ in their work).  $T_A$ is also the temperature below which the mode-coupling becomes appreciable~\cite{Gotze1992,Soklaski2016}.  The fundamental time-scale for viscosity is the Maxwell relaxation time, $\tau_M = \eta / G_{\infty}$, where $G_{\infty}$ is the infinite-frequency shear modulus.  Recent molecular dynamics (MD) studies of metallic liquids suggest that for $T > T_A$, $\tau_M$ is approximately equal to $\tau_{LC}$, the time required to change the coordination number of a local cluster by one~\cite{Iwashita2013,Iwashita2014,Soklaski2016}. 
	
	To evaluate the MD prediction inelastic neutron scattering measurements were made on liquid $\zp$.  To access the supercooled state and to avoid contamination all measurements were made with the liquid held in a containerless environment in high vacuum using the Neutron Electrostatic Levitation (NESL) facility~\cite{Mauro2016} located at the Spallation Neutron Source (SNS).  The results were converted into the time-dependent pair-distribution function, \emph{i.e.} the Van Hove function~\cite{VanHove1954}, $G\left(r,t\right)$, which allowed a study of the spatial and temporal correlations of the atoms. Due to experimental difficulties, studies of the Van Hove function in the past have been largely limited to computer simulations. Only in a few cases have measurements of $G\left(r,t\right)$ been made in metallic liquids at the melting temperature~\cite{Dahlborg1989}, and for water by inelastic x-ray scattering~\cite{Iwashita2017,Shinohara2018}. While $\tau_{LC}$ cannot be measured directly from experiment, new MD results discussed here show that it can be related to the decay time of the first peak area in the distinct part of the Van Hove function, $G_d\left(r,t\right)$. A comparison of the activation energies of $\tau_{VH}$ and $\tau_M$ confirms the prediction that both have an Arrhenius temperature dependence and the same activation energy. To our knowledge, this is the first significant experimental evidence indicating that local structural rearrangements underlie the dynamical behavior of high temperature metallic liquids.

\section{Experimental and Simulation Methods}\label{methods}
\subsection{Experimental Methods}
	Measurements of the high temperature properties of liquid metals such as Zr are often plagued by sample reactivity and oxygen contamination. These are minimized by processing the liquids without a container in a high vacuum environment using the technique of electrostatic levitation~\cite{Rhim1985}. The viscosity measurements were made with the Washington University Beamline Electrostatic Levitation~\cite{Mauro2011} (WU-BESL) facility; the experimental methods are discussed elsewhere ~\cite{Rhim1999,Gangopadhyay2017}. Inelastic neutron scattering measurements were made at Oak Ridge National Laboratory (ORNL) on the wide angular-range chopper spectrometer~\cite{Abernathy2012} (ARCS) beamline at the Spallation Neutron Source (SNS). The samples were processed in high vacuum using the Neutron-ESL (NESL) facility, which is optimized for both elastic and inelastic time-of-flight (TOF) neutron scattering studies~\cite{Mauro2016}.  The TOF inelastic neutron diffraction measurements on the levitated liquid samples were made with an incident energy $E_i$= 20meV. Due to the kinematic restrictions inherent to inelastic neutron scattering experiments, however, the maximum $q$ range for this incident energy is restricted to $q < 6\mathrm{\AA}^{-1}$. Though this restricted $q$-range can introduce termination ripples in the spatial Fourier transform to obtain the Van Hove correlation function, the increased energy resolution was deemed to be more important for the data needed.
	
	The samples studied were prepared from $\zp$ master ingots (1-2g, using high purity Zr (99.97\%) and Pt (99.997\%)), which were made by arc-melting on a water-cooled hearth in a high purity (99.999\%) Ar atmosphere. To further reduce the oxygen concentration in the atmosphere, a Ti-Zr getter was melted prior to arc-melting the ingot material. Both the ingot and the getter were held in the liquid state for $\sim$60 sec. This procedure was repeated three times, flipping the samples between melting to further increase mixing. The ingots were subsequently crushed and portions were used to create smaller spherical samples for viscosity ($\sim$45 mg) and inelastic neutron scattering measurements ($\sim$350 mg).
	
	To process the inelastic neutron scattering samples two fiber-coupled diode lasers (980nm,110W continuous maximum power output) were focused on opposite sides of the samples to reduce the temperature gradient. The sample temperature was measured using a single Process Sensors Metis MQ22 two-color ratio pyrometer. To obtain a sufficient scattering signal samples were held for $\sim$1.5-2hrs. at each temperature. Occasionally it was not possible to maintain levitation of the same sample for this full time, in which case a different sample was used to obtain the remaining data and the two scattering data sets were combined. The temperature data was corrected after processing by using the solidus temperature as a point of reference. A more detailed discussion of this technique can be found elsewhere ~\cite{Bendert2014}.
	
\subsection{Molecular Dynamics Simulations}
	Molecular Dynamics (MD) simulations were used to assist the data analysis. They were performed using the LAMMPS software~\cite{Plimpton1995a} with the $\mathrm{Zr}$-$\mathrm{Pt}$ embedded atom~\cite{Cheng2011} (EAM) potential developed by H. Sheng~\cite{Hirata2013}. The $\zp$ system was simulated with 15000 atoms under the NPT ensemble ($P=0$) with periodic boundary conditions. The Nos\'{e}-Hoover thermostat~\cite{Nose1984a,Hoover1985} was used to equilibrate the system at a target temperature for 15ns before data collection.  The Maxwell time was calculated from the atomic level stress using the Green-Kubo formula (shown below in Eq.~\ref{eq:GK}).
	
	These simulations were made using the high performance computing cluster in the Physics Department at Washington University in St. Louis. The atoms in the simulation were initialized to random locations. The system was then allowed to relax to remove overlapping atoms and evolve for 0.5ns at a high temperature to more closely resemble the liquid (\emph{i.e.} build local clustering and packing more akin to the inherent liquid structure). The simulation was cooled ($7 \times 10^{11}$ K/s) and subsequently equilibrated at each target temperature for 15ns. After these initial steps, data were collected for the structure (Van Hove correlation function ($G(r,t)$)), and the viscosity ($\eta$).
	
\subsubsection{Viscosity}
	As mentioned in the text the viscosity can be calculated using the Green-Kubo formula:
\begin{equation}\label{eq:GK}
\eta = \frac{V}{k_b T} \int \left\langle \sigma^{xy}(t)\sigma^{xy}(0) \right\rangle \mathrm{d}t \mathrm{ ,}
\end{equation}
where $V$ is the volume, $T$ is the temperature, and $\sigma^{xy}(t)$ is the shear stress at time $t$. The values obtained for this paper used the generalized Green-Kubo formula derived by Daivis and Evans \cite{Daivis1994}. The stress tensor was recorded for 4ns. at each temperature. The autocorrelation function for $\sigma^{xy}$ was computed using Fourier transforms according to the Weiner-Khinchin thorem:
\begin{equation}
\left\langle \sigma^{xy}(t)\sigma^{xy}(0)\right\rangle = \mbox{IFT}\left[ \mbox{FT}\left[\sigma^{xy}\right]\mbox{FT}\left[\sigma^{xy}\right]^{*}\right]
\end{equation}
where $\mbox{IFT}$ and $\mbox{FT}$ are the inverse and forward Fourier transforms and $^*$ indicates complex conjugation. From the viscosity the Maxwell time, $\tau_{M}$, was computed using $\tau_{M} = \eta/G_{\infty}$ where $G_{\infty}$ is the instantaneous shear modulus given by 
\begin{equation}
G_{\infty} = \frac{V}{k_b T} \left\langle \left(\sigma^{xy}\right)^2 \right\rangle
\end{equation}

\subsubsection{Van Hove Correlation Function}
	The distinct Van Hove correlation function is given, in MD simulations, by:
\begin{equation}
G_{d}(r,t) = \sum_{\alpha,\beta} \frac{c_{\alpha}c_{\beta}b_{\alpha}b_{\beta}}{\left[\sum_{\gamma} c_{\gamma}b_{\gamma} \right]} G_{d}^{\alpha\beta}(r,t) \mathrm{ ,}
\end{equation}
where $c_{\alpha}$ is the concentration and $b_{\alpha}$ is the scattering length of element $\alpha$, with $\alpha \beta$ ranging over all atomic pairs. $G_{d}^{\alpha\beta}(r,t)$ is the partial distinct Van Hove correlation function given by
\begin{equation}
G_{d}^{\alpha\beta}(r,t) = \frac{N}{\rho N_{\alpha}N_{\beta}}\sum_{i}^{N_{\alpha}}\sum_{i \neq j}^{N_{\beta}} \langle \delta(r-r_i(0)+r_{j}(t)) \rangle \mathrm{ ,}
\end{equation}
where $N_{\alpha}$ is the number of atoms of $\alpha$ and $\rho$ is the number density. Structural data used to calculate $G_d(r,t)$ were collected for 10ps at each desired temperature. The method for computing the Van Hove time is the same as for the experimental data which is discussed later. 

\section{Analysis, Results, and Discussion}

\begin{figure}
\centering
\includegraphics[width=\linewidth]{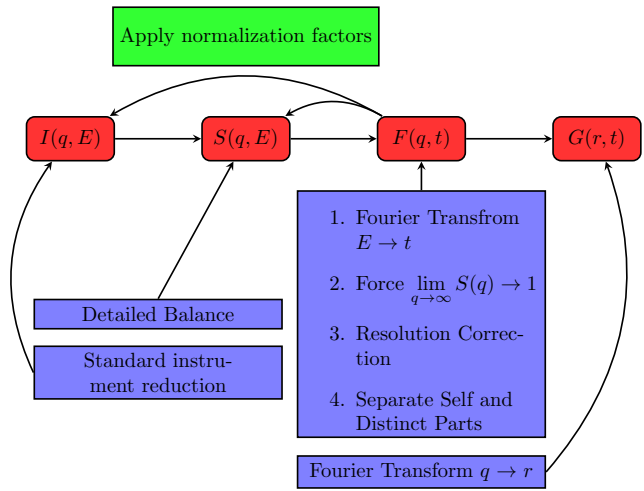}
\caption{\label{fig:FlowChart} A flowchart that describes a simplified data reduction method for inelastic neutron scattering measurements. The curved boxes indicate the functions while square boxes correspond to an analysis technique applied to the function or used to obtain the function.}
\end{figure}
	
\begin{figure}
\subfloat{\label{fig:SQE}}{\includegraphics[width=\linewidth]{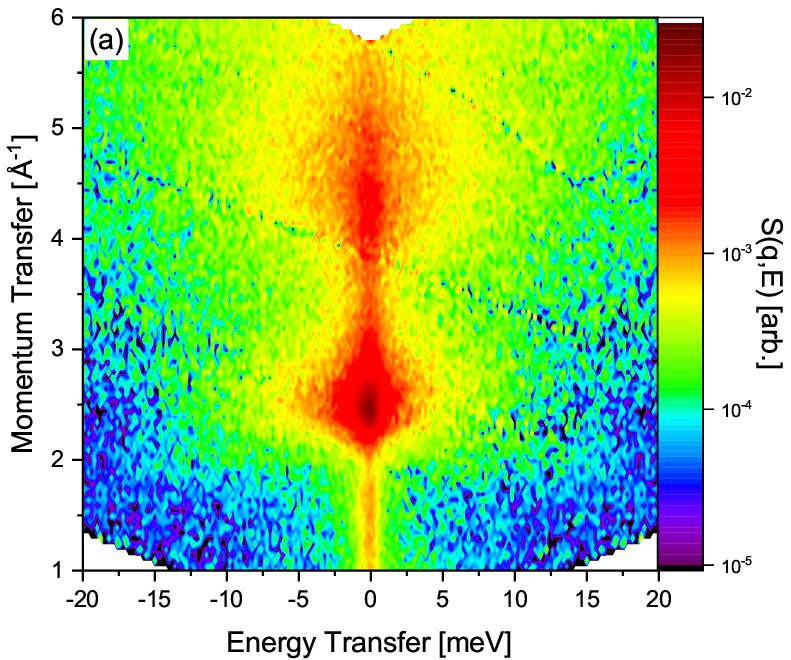}}
\subfloat{\label{fig:gdrt}}{\includegraphics[width=\linewidth]{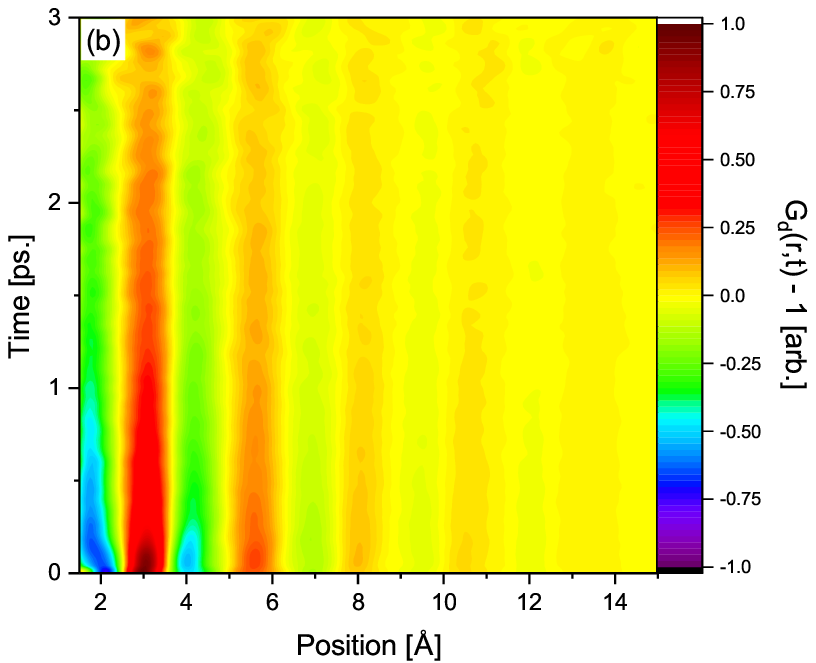}}
\vspace{-0.2in}
\caption{\label{fig:INSdata} Inelastic neutron scattering data for $\zp$ at $1833$K with $E_i = 20$meV. (a) The dynamic structure factor, $S(q,E)$ correcting for physical normalization ($S(q) \rightarrow 1$ as $q \rightarrow \infty$) and detailed balance. (b) The distinct Van Hove correlation function, $G_d(r,t)-1$, with the same corrections and the correction for the resolution function.}
\end{figure}

\begin{figure}
\centering
\includegraphics[width=\linewidth]{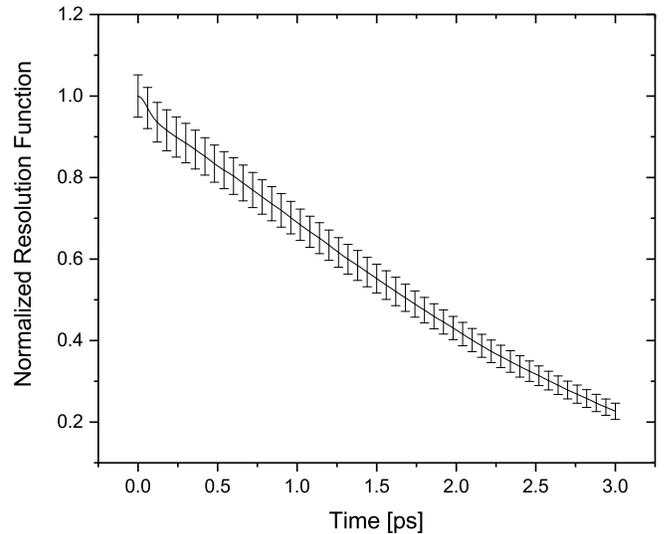}
\caption{\label{fig:ResFunc}Resolution function calculated from the Fourier transform of the vanadium normalized to its value at $t=0$ps. The value and error are computed from the mean and standard deviation, respectively, assuming that $F(q,t)$ is $q$-independent.}
\end{figure}

  The steps to obtain the dynamic structure factor $S(q,E)$ include the conversion of the TOF data to energy and momentum transfer, a physical normalization factor, the assumption of detailed balance and a correction for the resolution of the spectrometer. These steps and those used to obtain $G(r,t)$ are summarized in Fig.~\ref{fig:FlowChart}. The initial conversion used a standard reduction routine employing the MANTID~\cite{Arnold2014} software.  The required source beam parameters~\cite{Abernathy2015} for this reduction were obtained from previous calibration experiments, since they could not be measured in the NESL studies due to the presence of an internal beam-stop.  For normalization, the condition that $S(q) \rightarrow 1$ as $q \rightarrow \infty$ was enforced, using the $S(q)$ obtained from the intermediate scattering function, $F(q,t)$. This normalization was checked by comparing the $S(q)$ obtained here with one obtained from earlier neutron and x-ray diffraction data.   Detailed balance was used to extend the negative energy transfer data into regions that are inaccessible in the positive energy transfer region. A typical $S(q,E)$ obtained after these corrections is shown in Fig.~\ref{fig:SQE}. Since $F'(q,t) = F(q,t)R(t)$, where $F'$ is the measured and $F$ is the true intermediate scattering functions and $R$ is the resolution function in the time domain, the true intermediate scattering function can be obtained by dividing the measured signal by the resolution function. The resolution function was obtained from inelastic scattering measurements from polycrystalline vanadium at room temperature and was Fourier-transformed to the time domain. It is assumed that the scattering from the vanadium is completely incoherent so that the result is the neutron beam profile convoluted with the resolution of the detectors (\emph{i.e.} $S'(q,E) = S(q,E)\ast R(q,E)$). It is also assumed that the resolution function is independent of $q$, which should be approximately true for small energy transfer about $|E|<10$meV (see~\cite{Lin2016a} Fig.4 for an example vanadium scattering profile) or for a small enough $q$-range. To approximate the resolution function only the restricted section of the vanadium data $1.0<q<2.5 \mbox{\AA}^{-1}$ was used. The calculated resolution function given by the Fourier transform of the $I(q,E)$ data is shown in Fig.~\ref{fig:ResFunc}. Due to this treatment of the resolution function correction noise at long time is amplified (\emph{i.e.} from the division by the relatively small signal at long times) causing plateauing artifacts mentioned later (see Fig~\ref{fig:NtN0_comp}).

	The intermediate scattering function, $F(q,t)$, is obtained by a Fourier transform of $S(q,\omega)$
\begin{equation} \label{eq:fqt}
F(q,t) = \int_{-\infty}^{\infty} S(q,\omega)e^{i \omega t}\mathrm{d}\omega \mathrm{ .}
\end{equation}
The self ($F_s (q,t)$) and distinct ($F_d (q,t)$) parts of the intermediate scattering function, which describe single particle and collective density fluctuations respectively, are extracted by assuming that the self-part has a Gaussian form, \emph{i.e.} $F_s (q,t) = A(t) \exp(-w(t)q^2)$~~\cite{Dahlborg1989},  where the decay parameter, $w(t)$, and the amplitude, $A(t)$ are fitting parameters. The Gaussian approximation comes from expressions for the self-part of the Van Hove correlation function, $G_s (r,t)$, which has a Gaussian dependence in $r$ in both the hydrodynamic and free-particle limits~\cite{Scopigno2005a,Copley1975}.  For intermediate times, which are of interest here, the Gaussian approximation should still be a good approximation~\cite{Copley1975,Hansen2013}.  The distinct Van Hove correlation function is obtained from the Fourier transform of $F_d (q,t)$
\begin{equation}  \label{eq:grt}
G_d(r,t)-1 = \frac{1}{2 \pi^2 \rho}\int_{-\infty}^{\infty} F_d(q,t) \frac{q}{r} \sin(qr)\text{d}q \mathrm{ ,}
\end{equation}
where $\rho$ is the number density for the sample. A representative example of the $G_d(r,t)$ obtained from the data was shown in Fig.~\ref{fig:gdrt}. At $t=0$, $G_d(r,0)$ is equal to the equal-time (snapshot) pair-density function, $g(r)$. The integrated peak intensity is computed for each temperature from

\begin{figure}
\includegraphics[width=\linewidth]{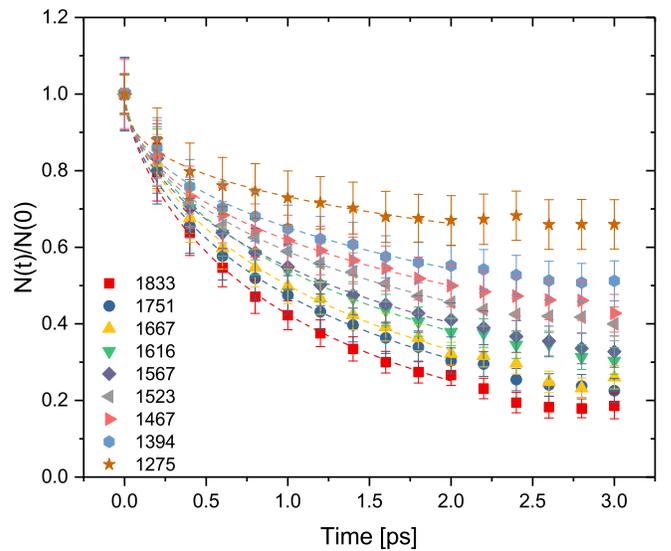}
\caption{\label{fig:NtN0} The normalized integrated peak intensity $N(t)/N(0)$ versus time plotted for each temperature (K).  The data is fit (dashed lines) out to 2.0 ps. using the stretched-exponential function (Eq.\ref{eq:KWW}).}
\end{figure}

\begin{equation}  \label{eq:Nt}
N(t) = \int_{D} 4\pi r^2 \rho (G_d(r,t)-1) \text{d}r \mathrm{ ,}
\end{equation}
where $D$ is the positive region of the first intense peak of the integrand. Because $G_d(r,t)$ decays to unity at large $t$, $G_d(r,t)=1$ provides the baseline to define the density fluctuation. $N(t)$ is proportional to the dynamic coordination number, and reflects the average decorrelation time for atoms located near the first peak of $G_{d}(r,t)$. The decorrelation time is the time for an atom initially located near the central atom to begin to diffuse away.  This time is a function of the distance that an atom is from the central atom’s initial position.  It also depends on the local structure, which can be quite varied~\cite{Johnson2016a}.  Since the exponential decay time for each atom is different, the overall decay of $N(t)$ could be described by a Kohlrausch-Williams-Watts (KWW) stretched exponential function~\cite{Cavagna2009},
\begin{equation} \label{eq:KWW}
y(t)= \exp{\left(-(t/\tau)^{\beta}\right)} \mathrm{ ,}
\end{equation}
where $\tau$ is the time constant and $\beta$ is the stretching factor.  From MD simulations and a recent study on water~\cite{Shinohara2018} $N(t)$ is expected to have two decay rates; one is due to ballistic motion and another that describes the changes in the configuration of the nearest-neighbors, \emph{i.e.} the opening of the cage. However, due to the limited energy range of these experiments for $\zp$ it was not possible to determine the decay rate in the ballistic region. The normalized peak intensity $N(t)/N(0)$ is shown in Fig.~\ref{fig:NtN0}. The initial decay in $N(t)/N(0)$, $t<0.1$ps, is due to ballistic motion and is only weakly dependent on temperature. As $t\rightarrow \infty$ it is expected that $N(t)/N(0) \rightarrow 0$ as the correlations between the initial position are lost. However, the data show plateaus at longer times. These are artifacts from the resolution function correction, as previously discussed, and are not fit to the KWW expression. Also, the ballistic region is not well described by the single KWW expression. The dashed lines in Fig.~\ref{fig:NtN0} are fits to the KWW expression, which describes well the data beyond $0.1\mathrm{ps}$ for all temperatures studied.

\begin{figure}
\centering
\includegraphics[width=\linewidth]{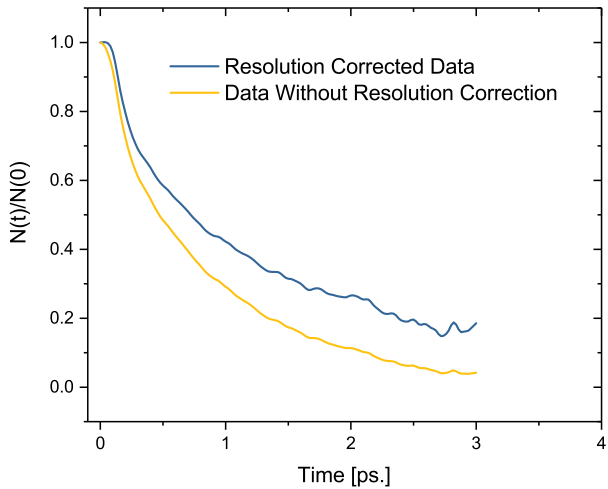}
\caption{\label{fig:NtN0_comp} Comparison of the normalized decay function $N(t)/N(0)$ obtained from the INS experiments at $T=1833\mathrm{K}$ using the resolution function correction (top curve) and without the resolution correction (bottom curve).}
\end{figure}

Figure ~\ref{fig:NtN0_comp} shows a comparison between the data for $N(t)/N(0)$ obtained from the inelastic neutron scattering experiments, both with the resolution function correction and without the resolution function correction. A noticeable effect coming from the resolution function correction is an elongation of the long time tail of the decay (\emph{i.e.} the plateauing mentioned previously). This is an artifact, arising from an incomplete knowledge of the true resolution function for the beam line. To account for this effect the decay data is only fit with Eq.~\ref{eq:KWW} out to $2\mathrm{ps}$ corresponding to when the resolution function decays to approximately $1/e$ of its initial value. The minimum time, set by the maximum energy transfer $20\mathrm{meV}$, is $\sim 0.2\mathrm{ps}$

\begin{figure}
\includegraphics[width=\linewidth]{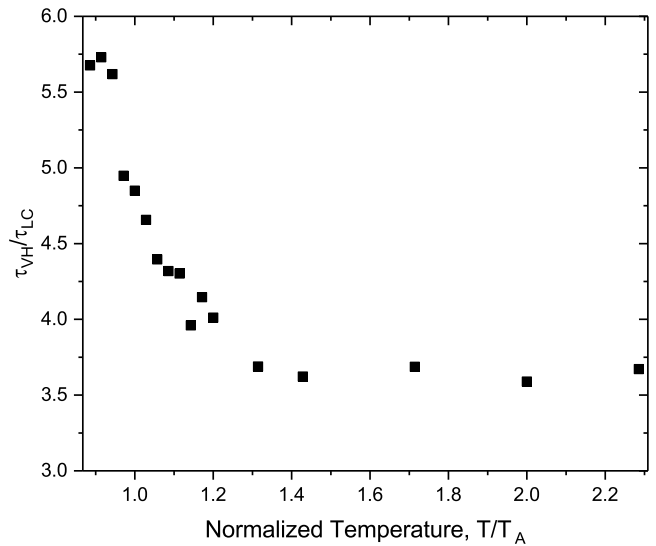}
\caption{\label{fig:Ratio} The ratio of the Van Hove time, $\tvh$, to the local configuration time, $\tlc$, as a function of temperature normalized to $T_A \approx 1750$K from MD simulations of liquid $\zp$. $\tvh / \tlc \approx 3.6$ for $T>T_A$.}
\end{figure}

	As mentioned earlier, the local configuration time, $\tlc$ cannot be obtained directly from the experimental data.   However, our MD simulations show that $\tlc$ is related to a measurable quantity called here the Van Hove time, $\tvh$, which is the long decay time corresponding to the configuration of nearest-neighbors in the first peak in $G_d(r,t)$.  The experimental value of the Van Hove time was obtained from the mean relaxation time of the KWW function fit to the data, $\langle \tau_r \rangle = \frac{\tau}{\beta} \Gamma( \frac{1}{\beta}) \equiv \tvh$. The results from the MD simulations shown in Fig.~\ref{fig:Ratio} indicate that $\tlc \approx \tvh/3.6$ for $T>T_A$ ($T_A \approx 1750\mathrm{K}$).  Since the ratio is approximately constant for $T/T_A > 1.2$ the activation energy for $\tvh$ will be the same as for  $\tlc$. For water $\tau_{LC}$ was approximately equal to $\tau_{VH}$ ($\tau_{2}$ in their work)~\cite{Shinohara2018}. Because metallic liquids have more nearest neighbors than water does ($\sim 13$ for metallic liquids and 4 for water), the ratio $\tau_{VH}/\tau_{LC}$ should reflect this difference~\cite{Wu2018}. The rise of the ratio $\frac{\tvh}{\tlc}$ below $\sim 1.2 T_A$, however, is small compared to the change in $\tvh$ with temperature.

\begin{figure}
\subfloat{\label{fig:TVH_Data}}{\includegraphics[width=\linewidth]{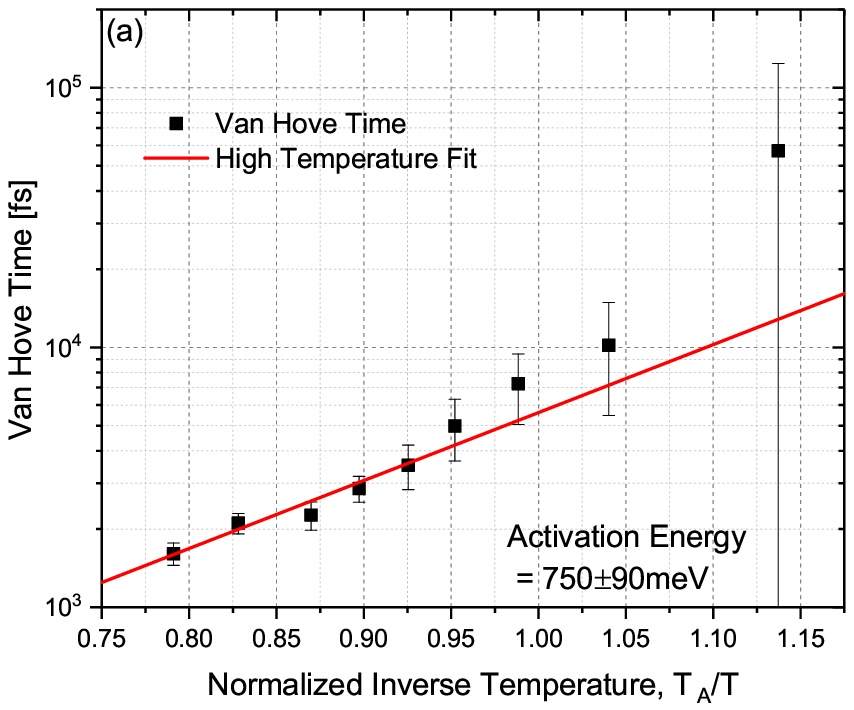}}
\subfloat{\label{fig:Visc_Data}}{\includegraphics[width=\linewidth]{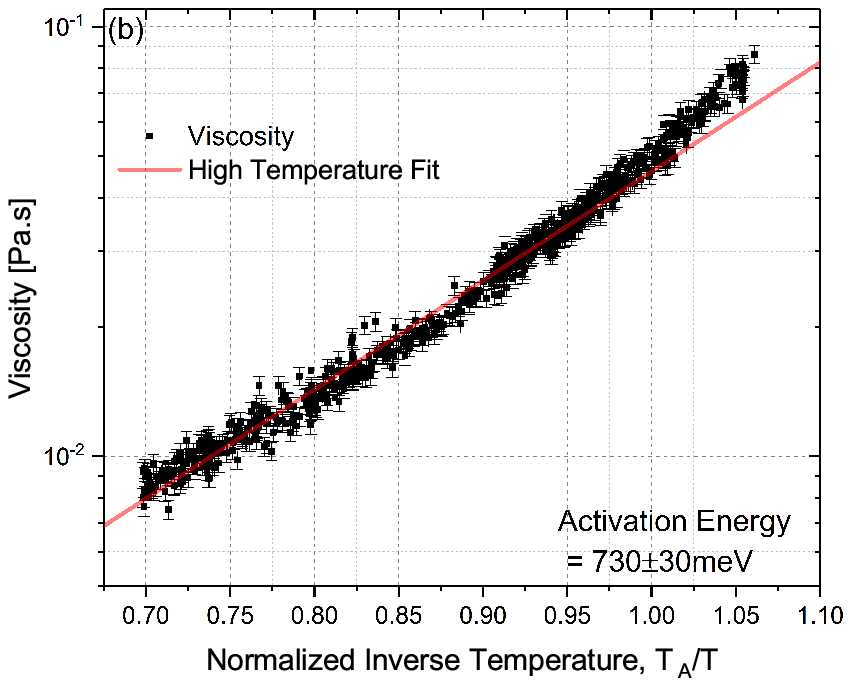}}
\vspace{-0.2in}
\caption{\label{fig:EAdata} (Color online) (a) Van Hove time and (b) viscosity data for liquid $\zp$ versus inverse temperature normalized to the Arrhenius crossover temperature determined from the viscosity ($T_A=1450$K). The best fit lines for the activation energy are also shown; the slopes give an activation energy of 730meV for the viscosity and 750meV for $\tvh$. The error bars shown for $\tvh$ are $3\sigma$ and are calculated from the error in the fit parameters from Eq.~\ref{eq:KWW}.}
\end{figure}

	As shown in Fig.~\ref{fig:TVH_Data}, $\tvh$ obtained from the scattering data shows an Arrhenius temperature dependence for  $T/T_A>1$, as indicated by the previous MD simulations for the related local configuration time, $\tlc$~\cite{Iwashita2014}. The results from the MD studies indicate that $\tlc$ remains Arrhenius far below $T_A$. Based on the results in Fig.~\ref{fig:Ratio}, $\tvh$ should become super-Arrhenius below this temperature, as suggested by the data in Fig~\ref{fig:TVH_Data}.  The activation energy for $\tvh$ (and from Fig.~\ref{fig:Ratio}, for  $\tlc$ ) above $T_A$ is 750$\pm$90meV.  As shown in Fig.~\ref{fig:Visc_Data}, the activation energy for the measured viscosity above $T_A$ is 730$\pm$30meV.  Within experimental error, then, these activation energies are equal, indicating that the energy barrier is the same for both processes. This provides experimental evidence that the MD predictions~\cite{Iwashita2013,Iwashita2014} are correct, \emph{i.e.} showing that the atomic rearrangements that determine $\tvh$ (and $\tlc$) are controlling the viscosity at high temperatures.  

\section{Conclusion}
	In summary, the time dependent pair distribution function (distinct Van Hove function, $G_d\left(r,t\right)$) was obtained as a function of temperature from inelastic neutron scattering data for equilibrium and supercooled $\zp$ liquids made in a containerless environment.   Molecular dynamics simulations showed that the relaxation time of the positive peak area in $G_d\left(r,t\right)$ (defined as the Van Hove time, $\tau_{VH}$ ) is related to the local configuration time, $\tau_{LC}$ , and has the same temperature dependence above the crossover temperature, $T_A$.   A comparison of the experimental neutron scattering and viscosity data show that the activation energy of $\tau_{VH}$ and that of the Maxwell time, $\tau_M$, are equal to within experimental error, strongly suggesting that they are governed by the same process . To our knowledge this is the first experimental evidence for a key prediction from recent MD studies for metallic liquids, which indicate that local structural excitations underlie the viscosity at high temperature.   
	
\section*{Acknowledgments}
	We thank A.J. Vogt, M.L. Johnson, C.E. Pueblo, R. Dai, H. Wang and W. Dmowski for their help in the Neutron scattering experiments and for useful discussion. The work at Washington University in St. Louis was partially supported by the National Science Foundation under Grant DMR-15-06553 and the National Aeronautics Space Administration (NASA) under grant NNX16AB52G. T. Egami and Z. Wang were supported by the U. S. Department of Energy, Office of Science, Basic Energy Sciences, Materials Science and Engineering Division. A portion of this research used resources at the Spallation neutron Source a DOE Office of Science User Facility operated by the Oak Ridge National laboratory. The work done at Iowa State was supported by the National Science Foundation under Grant No. DMR-1308099.
\bibliographystyle{apsrev4-1}
\bibliography{library}

\end{document}